\documentclass[]{spie}
\usepackage[]{graphicx}

\def\microas{\mu{\rm as}}
\def\Msun{\rm{M_\odot}}

\def\paramPn{5.9691}
\def\errPn{0.0022}

\def\paramin{119.7}
\def\errin{2.8}

\def\paraman{864}
\def\erran{27}

\title{Initial Scientific Results from Phase-Referenced Astrometry of Sub-Arcsecond Binaries}

\author{Matthew W. Muterspaugh\supit{a}\\
Benjamin F. Lane\supit{a}\\
B.~F.~Burke\supit{a}\\
Maciej Konacki\supit{b}\\
S.~R.~Kulkarni\supit{c}
\skiplinehalf
\supit{a}Massachusetts Institute of Technology, Center for Space Research, 70 Vassar Street, Cambridge, MA, U.S.A.; \\
\supit{b}California Institute of Technology, Div.\ of
Geological \& Planetary Sciences 150-21, Pasadena, CA 91125, U.S.A.;\\
\supit{c}California Institute of Technology, Astronomy Department,  Mail Code 105-24, 
1200 East California Blvd, Pasadena, CA 91125 U.S.A.
}

\authorinfo{\noindent Send correspondence to M.W.M.\\
M.W.M.: E-mail: matthew1@mit.edu, Telephone: 1 617 258 7805\\
M.K.: E-mail: maciej@gps.caltech.edu}

\begin{document} 
\maketitle

\noindent
{\bf Copyright 2004 Society of Photo-Optical Instrumentation Engineers.\\}
This paper will be published in SPIE conference proceedings volume 5491, 
``New Frontiers in Stellar Interferometery.''  and is made available as 
and electronic preprint with permission of SPIE.  One print or electronic 
copy may be made for personal use only.  Systematic or multiple reproduction, 
distribution to multiple locations via electronic or other means, duplication 
of any material in this paper for a fee or for commercial purposes, or 
modification of the content of the paper are prohibited.

\begin{abstract}
The Palomar Testbed Interferometer has observed several binary star
systems whose separations fall between the interferometric coherence
length (a few hundredths of an arcsecond) and the typical atmospheric
seeing limit of one arcsecond.  Using phase-referencing techniques we
measure the relative separations of the systems to precisions of a few
tens of micro-arcseconds.  We present the first scientific results of
these observations, including the astrometric detection of the faint third
stellar component of the $\kappa$ Pegasi system.
\end{abstract}

\keywords{Optical Interferometry, Phase-referencing, astrometry, PTI, binary star, kappa Pegasi, kappa Peg}

\section{Introduction}
\label{sect:intro}  
A new method of ground-based differential astrometry with measurement precisions 
on the order of $10^{-5}$ arcseconds for bright stars separated by 
$\approx 0.05-1.0$ arcseconds has been developed for use 
at the Palomar Testbed Interferometer \cite{LaneMute2004a}.  Observations 
using this method have been carried out 
over the past year on 25 binary systems as 
part of the Palomar High-precision AStrometric Exoplanet Search (PHASES).  
PHASES will monitor up to 50 such systems 
over $\approx 3$ years for a number of scientific 
purposes, including a search for astrometric 
perturbations caused by faint (planetary) companions 
orbiting one of the main stars of a system 
(``S-type orbits,'' as opposed to ``P-Type'' 
companions which orbit both stars).

One system in the PHASES study is $\kappa$ Pegasi (HR 8315, HD 206901).  
This system is comprised of two 
F5 subgiant stars (here referred to as A and B; 
for historical reasons, B is the 
brighter and more massive---this distinction has 
been the cause of much confusion) and 
a faint stellar companion orbiting B 
(which this paper designates as ``b'').  
An additional component C is well 
separated from the other members 
of the system (13.8 arcseconds) 
and is faint; this may be optical and 
is not relevant to the present analysis.
Using the best available values for the 
masses of stars A and B, combined with the 
most recent elements of the A-B 
visual orbit \cite{Soder1999}, the stability 
criteria of Holman and Wiegert \cite{holman1998} 
for S-type companions predicts that objects with 
orbital periods as long as five months are stable in this system.
Both A and B have been reported as suspected 
spectroscopic subsystems (here referred to as 
A-a and B-b, respectively), but no companion 
to A has been confirmed.

Burnham discovered the sub-arcsecond 
A-B binary in 1880 \cite{Burnham1880}.  
Since this discovery, an number of 
studies have been carried out to determine 
the orbit of A-B and to search 
for additional components.
In 1900 Campbell \& Wright \cite{CampbellWright1900} 
reported a period and semimajor axis 
for the A-B pair of 11 years and 0.4 arcseconds, 
respectively, and that the brighter 
of the stars is a spectroscopic binary 
with a period ``that seems to be about six days.''
Luyten \cite{Luyten1934} combined all 
previous observational data to produce a visual 
orbit between components A and B and a  
spectroscopic orbit for the 5.97 day B-b system 
(he interchanged the designations A and B; here we have 
converted his results to the convention previously mentioned).  
His work also discredited previous 
claims that the line of apsides of B-b 
varied with the period of the A-B system.  
Luyten derived a mass for component A of 1.9 $\Msun$ and a
combined mass for the B-b subsystem of 3.3 $\Msun$.  
Additionally, because there are no observed 
eclipses in the B-b system, he concluded that the 
maximum possible mass ratio $M_B$:$M_b$ is 3:1.
Beardsley \& King \cite{BeardsleyKing1976} obtained 
separate spectra for components A and B.
Their observations confirmed that component B is a 
5.97 day single-line spectroscopic binary, and also 
suggested that A was a spectroscopic binary with period 4.77 days.  
Mayor \& Mazeh \cite{MayorMazeh1987} have published the most recent 
spectroscopic orbit for the B-b subsystem, 
as well as several measurements of the radial 
velocity of component A, which did not confirm the 
proposed 4.77 day companion a.  Mayor \& Mazeh appear to switch naming 
conventions for components A and B several times in their paper.  They report 
a mass ratio ``$M_A$:$M_B = 1.94\pm0.6$''; this is counter to the tradition 
of $\kappa$ Pegasi B being the more massive star, and the 
correct value is probably the inverse of this.  Mayor \& Mazeh indicate 
that it is component B that contains the 5.97 day spectroscopic binary.
The most recent visual orbit for system A-Bb was 
published by S\"oderhjelm using historical data 
combined with Hipparcos astrometry\cite{Soder1999}.  
Because the period is short and Hipparcos was capable 
of wide-field astrometry, estimates for the 
parallax ($27.24\pm0.74$ mas), total mass (4.90 $\Msun$), 
and mass ratio of components A and B ($M_B$:$M_A=1.76\pm0.11$, 
in inverse agreement with Mayor \& Mazeh) were also possible.  
This paper reports astrometric PHASES data that detects 
the reflex motion of the center-of-light of the 
B-b subsystem relative to the primary star A, with a period of 5.97 days.

Observations were made using the Palomar Testbed Interferometer (PTI)\cite{col99}.  
PTI is located on Palomar Mountain near San Diego, CA. It was developed by
the Jet Propulsion Laboratory and California Institute of Technology for
NASA as a testbed for interferometric techniques applicable to the
Keck Interferometer and other missions such as the Space
Interferometry Mission, SIM.  It operates in the J ($1.2 \mu{\rm
m}$), H ($1.6 \mu{\rm m}$), and K ($2.2 \mu{\rm m}$) bands, and combines
starlight from two out of three available 40-cm apertures. The
apertures form a triangle with 86 and 110 meter baselines.  The differential astrometry 
mode used by the PHASES project has been demonstrated using K band light 
from the North-South (110 meter) and the South-West (86 meter) 
baselines; the North-West (86 meter) baseline will become 
operational for this mode this year.  All data presented in 
the current paper were taken using the North-South (longest) baseline.

\section{Observational Setup and Data Reduction Process}
The observational setup used is described in detail in 
Lane \& Muterspaugh 2004\cite{LaneMute2004a}.  The technique 
is built upon the fact that, in an interferometer, 
the position of zero total optical path delay 
(the point at which broadband light interferes 
with maximum intensity) 
depends on the direction to the light source.  
By measuring the relative zero-delay locations of stars, 
one can determine their relative sky positions.  
If the stars are separated by a small angle on the sky, 
atmospheric contributions to the optical delay are 
common and cancel.  The atmospheric terms in the 
optical delay are not constant in time; in the 
absence of making simultaneous measurements of all stars, 
one must be able to monitor and correct for 
changes in the atmospheric optical delay.  We make 
this correction using a technique known as phase-referencing.

Each of two telescopes collects light from all stars in 
a target field of approximately 1 arcsecond.  
The light is then passed through low-vacuum pipes 
to an optical lab where movable mirrors (delay lines) 
introduce optical delays to compensate for
the geometric delays associated with an interferometer.  
The light from all stars go through a common path 
through the delay lines; no significant amount of differential 
delay is added to the light of various stars in the 
field through this process.  The light from each telescope 
is then split using beamsplitters.  
A portion of this light (roughly 70\%) from each telescope 
is combined to produce an interference pattern on a 
high-speed (10-ms) fringe-tracking detector.  
This actively monitors the fringe position of
one star in the field to determine the delay 
motion added by the atmosphere.  This measurement 
is fed back to the optical system to remove a 
portion of the temporal effects of the atmosphere.

The rest of the starlight (30\%) is combined separately.  
This ``science'' combiner is able to take measurements 
over much longer time periods because it is phase-referenced to the 
fringe-tracking combiner, which removes much of the 
atmospheric delay variability.  For this experiment, 
we choose to scan in delay with relatively large amplitude 
(on order 100 wavelengths, or 220 microns) so as to 
measure the interferograms at the locations of each star 
in the system.  The scanning is done in a triangle 
waveform with periods on the order of 2-3 seconds.
A laser metrology system is used to measure the 
differential path length between the fringe-tracking 
and science beam combiners.

The data reduction process was similar to that 
described in Lane \& Muterspaugh, 2004.  
The science-combiner data is broken into ``scans'' each 
time the scanning delay direction changes.  A likelihood 
function of the projected delay separation  is determined 
on a scan-by-scan basis by creating a model double fringe 
packet for all possible delay separations and 
evaluating $\chi^2$ for each at all possible delay positions; 
$\chi^2$ is used as the figure of merit to construct the likelihood function.    
Because there are multiple fringes in each star's interferogram, 
this likelihood function has many local maxima and 
minima separated by the fringe spacing.  For a typical target, 
the signal-to-noise ratio in a single scan is not large enough to determine 
the correct global maxima, resulting in a periodic 
ambiguity in the projected separation.  The likelihood function 
is remapped into a grid of differential declination and 
differential right ascension.  The likelihood functions 
for all scans (typically 500-3000 scans per target per night) are coadded in 
this grid, improving the SNR enough to determine 
the sky separation of the two stars observed unambiguously.  
A direct fit of a rotated two-dimensional quadratic to the 
coadded likelihood function determines the formal 
uncertainties of the astrometric measurement.  
The two-dimensional likelihood function 
itself is used to plot error ellipses.

The data analysis presented in this paper does 
not include the effects of aberration, precession, 
nutation, or parallax.  These astrometric corrections 
are predicted to affect the results in two ways.  
The most obvious effect is that the apparent global astrometry 
for the system as a whole will be different than that 
used.  An error of 1 arcsecond in mean global system 
astrometry introduces an error in the measured 
separation of $10^{-6}$ arcseconds.  These astrometric 
corrections can be as large as a few tens of arcseconds, 
resulting in differential astrometry errors of magnitude 
similar to our observed precisions ($10^{-5}$ arcseconds).  
A second effect is due to a variation in true North with time.  
This will introduce errors in measured position angles.  
Because the observations presented were carried out over a 
timescale of a few months and these effects are slowly varying, 
the precision of the data and observational repeatability will not be 
affected in any significant manner, nor will the precisions 
with which orbital elements are able to be determined from the data.  
To determine orbital elements with accuracies equal to the 
observed measurement precision will require including these 
astrometric terms; to the extent that these effects are significant, 
all orbits presented in this paper are preliminary.  
However, the size of the reflex motion of $\kappa$ Pegasi B 
by the unseen companion b is an order of magnitude larger 
than the expected effect any of these corrections will have on
the differential astrometry measurements.  
We are working to include these astrometric corrections 
for future analysis.

\section{Observational Performance}
PHASES has observed 25 binary systems each multiple times.  
The data from a number of these has been reduced and analyzed to 
determine observational repeatability.  See ``Phase Referencing 
and Narrow-Angle Astrometry in Current and Future Interferometers,'' 
by Lane \& Muterspaugh in this proceedings, for more details.

Most of the $\kappa$ Pegasi data presented in this paper was 
taken during the 2003 observing season.  Over a similar time range 
the long-period (order century) systems HR 6983 (HD 171779) and 
72 Pegasi (HR 8943, HD 221673) were observed in the same mode, each 
with over 20 nights of observations.  For both systems the formal 
uncertainty error ellipses were typically $5\times100$ $\microas$ 
(1 $\microas=10^{-6}$ arcseconds);
for a few data points taken over a wide range of hour angles, the major 
axis uncertainties were nearly equal to those of the minor axis.  
The orbital motions of these long period systems are well described 
by linear or polynomial models; a fit of this type establishes our 
night-to-night repeatability.  For HR 6983, the reduced  
$\sqrt{\chi_r^2} = 4.7$, and a similar factor is found for 72 Pegasi.    
Separate fits to the one-dimensional uncertainties in the $\delta{\rm R.A.}$
or $\delta{\rm Dec}$ axis produce similar values, indicating that the 
scale factor between formal and actual errors is uniform.  Our 
demonstrated night-to-night repeatability is 24 $\microas$ for these systems.

To reduce this scaling factor, we have added a laser metrology system 
to a short amount of differential starlight path that had previously 
been unmonitored; this improved system is in use for the 2004 observing season.  
Initial repeatability measurements on the star $\beta$ CrB (HR 5747, HD 137909) 
indicate that this scale factor has been reduced to approximately 2.6.  
A direct comparison using the same targets must wait until summer 2004, when 
HR 6983 and 72 Pegasi are again observable.

\begin{figure}[h]
\centerline{\hspace{1.4in}\includegraphics[height = 4.80 in, angle = 270]{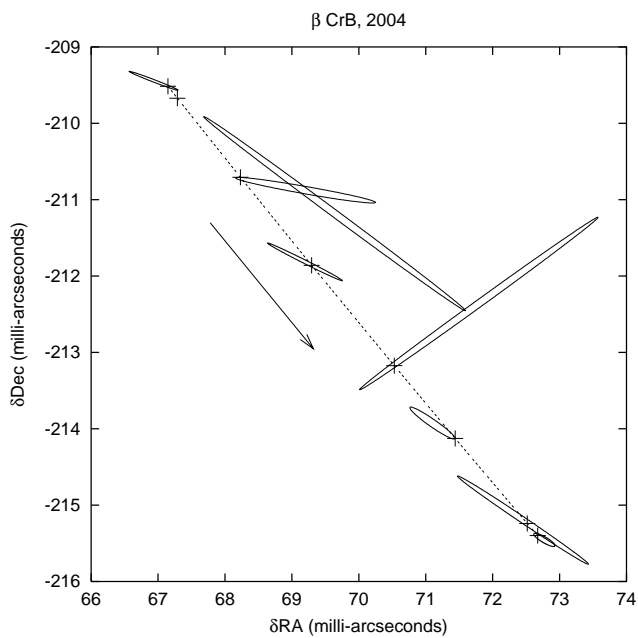}}
\caption{Observed repeatability of the differential astrometry of $\beta$ CrB during the 2004 
observing season.  Error ellipses have been rescaled by the appropriate factor (2.6).  
Arrow indicates direction and rate of motion, with length 10 days.}
\end{figure}

The appropriate scaling factor has been applied to the formal uncertainties 
of $\kappa$ Pegasi data presented in this paper; the correction has been applied 
to the error bars and ellipses plotted in all figures.  Due to the orientation of 
the baseline with the binary's separation angle, the target was typically 
observable for much less than an hour each night.  We find the average scaled 
minor-axis uncertainty for $\kappa$ Pegasi is 45 $\microas$, and that for 
the major axis is 1.5 mas.  This extreme axis-ratio is to be expected 
from a single baseline interferometer.  It should be noted, however, that 9 data 
points have scaled minor-axis uncertainties of less than 30 $\microas$ and 8 have 
scaled major-axis uncertainties of less than 250 $\microas$.  All data points have 
scaled minor-axis uncertainties less than 1/8 of the observed semi-major axis of the 
orbit of the B-b center-of-light.

\section{Astrometric Model}
We have applied basic models to our astrometric data.  
We make the simplifying assumption that the B-b subsystem is 
unperturbed by star A over the timescale of our observing program, 
allowing the model to be split into a wide (slow) interaction 
between star A and the center of mass of Bb, and the 
close (fast) interaction between stars B and b.  
The preliminary results presented in this paper result from modeling 
both the A-Bb and B-b motions with Keplerian orbits.  Because our data 
were taken over a time short compared to the period of the A-Bb system, 
the period and eccentricity of the wide system were held fixed at 
the values determined by S\"oderhjelm.

In general, one cannot simply superimpose the results of the two orbits.  
The observable in our measurements is the separation of star A 
and the center-of-light of the B-b subsystem.  
Because the center-of-light of B-b, the center-of-mass of B-b, 
and the location of star B are generally all 
unequal, a coupling amplitude must be added to the combined model.  
This coupling amplitude measures the relative size of the semi-major axis 
of the B-b subsystem to that of the motion of the center-of-light 
of the B-b subsystem.  The sign of the superposition is determined by 
the relative sizes of the mass and luminosity ratios of the stars B and b.  
As an example, if the center-of-light is located between the center-of-mass of B-b 
and the location of star B, the motion of the center-of-light will be 
in opposite direction to the vector pointing from B to b.
For a subsystem with mass ratio 
$M_{\rm{b}}/M_{\rm{B}}$ and luminosity ratio $L_{\rm{b}}/L_{\rm{B}}$, the observed quantity is
$$
\overrightarrow{y_{\rm{obs}}} = \overrightarrow{r_{\rm{A-Bb}}} - \frac{M_{\rm{b}}/M_{\rm{B}} - L_{\rm{b}}/L_{\rm{B}}}{\left(1+M_{\rm{b}}/M_{\rm{B}}\right)\left(1+L_{\rm{b}}/L_{\rm{B}}\right)}\overrightarrow{r_{\rm{B-b}}}
$$
where $\overrightarrow{r_{\rm{A-Bb}}}$ is the model separation pointing from star 
A to the center-of-mass of Bb, and $\overrightarrow{r_{B-b}}$
is the model separation pointing from star B to star b.  Including this coupling term for astrometric data is 
important when a full analysis including radial velocity data is made.

Alternatively, one can directly combine a model of the A-Bb system 
with a model of the motion of the center-of-light of B-b.  
For purely astrometric data such a model is appropriate.  In this case, 
there is no sign change for the B-b center-of-light model, 
and no extra coupling amplitude is required.  This was the model used 
to construct preliminary orbits for the current paper.
$$
\overrightarrow{y_{\rm{obs}}} = \overrightarrow{r_{\rm{A-Bb}}} + \overrightarrow{r_{\rm{Bb, C.O.L.}}}
$$

Radial velocity data are available for $\kappa$ 
Pegasi A and B (e.g. Mayor \& Mazeh 1987).  
An effort to incorporate these data in a full three-dimensional 
double-orbit model is currently under way.

\section{Analysis and Discussion}

The data was simultaneously fit to two Keplerian models; one representing 
the A-Bb interaction, and another representing 
the motion of the center-of-light of the B-b subsystem.  The period and 
eccentricity of the wide system were fixed at 
S\"oderhjelm's values of 4233 days and 0.31, respectively; 
our few months of observation do not allow these parameters to be well-determined.  
All elements of the preliminary A-Bb orbit agree 
well with both Luyten's and S\"oderhjelm's results.

\begin{figure}[!h]
\centerline{\hspace{1.4in}\includegraphics[height = 5.3 in, angle = 270]{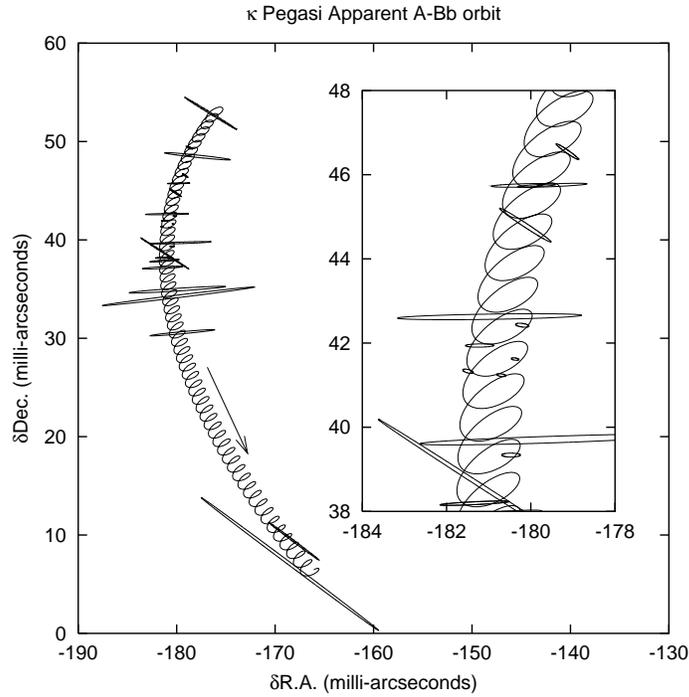}}
\caption{A preliminary apparent orbit between $\kappa$ Pegasi A and the center-of-light of 
subsytem Bb, over the time period of PHASES data.  Arrow indicates direction of increasing time, 
with a length of two months.
Also plotted are the 1-$\sigma$ error ellipses for PHASES data.  The error ellipses are generally 
much longer in one dimension than the other because a single-baseline interferometer is more 
precise in measuring quantities parallel to the baseline vector.}
\end{figure}

PHASES data is particularly well-suited to determining the orbital elements of the B-b subsystem.  
Because the B-b subsystem has short period (our fit 
gives $\paramPn\pm\errPn$ days), the system can be studied in a 
relatively short time.  The apparent semi-major axis of the center-of-light 
of B-b is roughly $\paraman\pm\erran$ $\microas$, 
a factor roughly 20 times larger than our average 
precision; this large signal makes $\kappa$ Pegasi Bb ideal for 
developing astrometric methods.
A full model of the A-Bb orbit will require either several years of
observation, or including historical micrometer or speckle interferometry data.

\begin{figure}[!h]
\centerline{\hspace{1.0in}\includegraphics[height = 4.9 in, angle = 270]{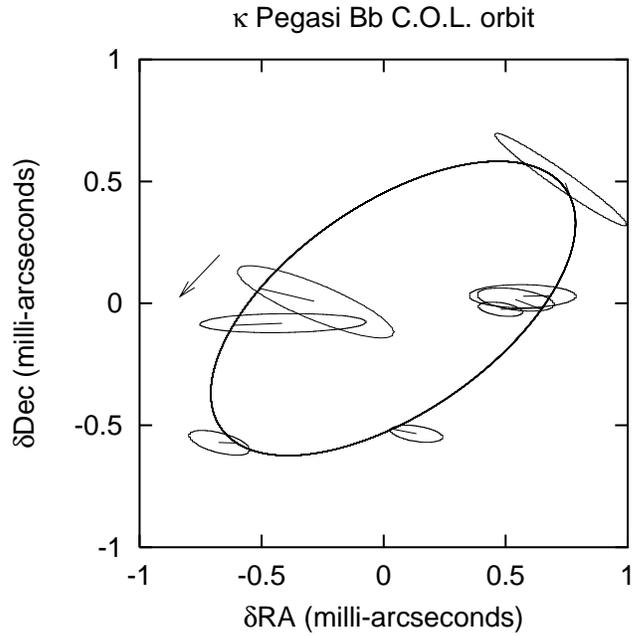}}
\caption{Preliminary orbit of the center-of-light of subsystem B-b, with the long-period A-Bb orbit removed.  
For clarity, only those data with both error 
ellipse axis smaller than the semi-major axis of the orbit have been plotted.
Lines connect each data 
point to its corresponding point in the orbit.  Arrow indicates orbital motion over 6 hours near periastron.}
\end{figure}

\begin{figure}[!h]
\centerline{\includegraphics[height = 3.75 in, angle = 270]{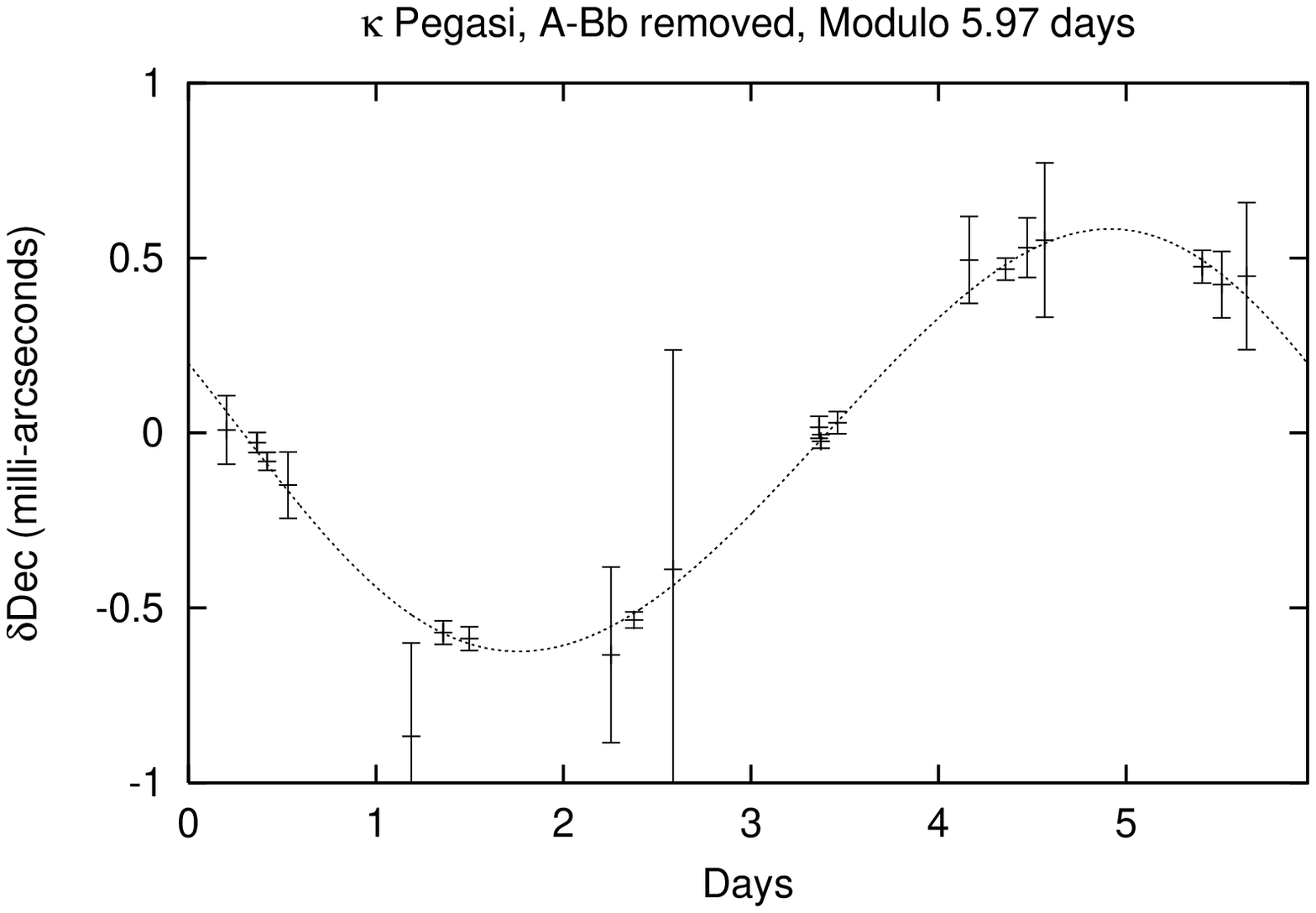}}
\caption{Residuals to the A-Bb model for $\kappa$ Pegasi, in differential declination 
(typically our more sensitive dimension), wrapped about the period of B-b ($\paramPn$ days).  
Also plotted is the preliminary Keplerian model of the motion of the center-of-light of B-b.
For clarity, 6 data points with declination uncertainties larger than the 
semi-major axis motion of the center-of-light of B-b 
($\paraman\times10^{-6}$ arcseconds) have not been included in the plot.}
\end{figure}

The period of the preliminary astrometric B-b orbit agrees well with the previous 
spectroscopic results ($\paramPn\pm\errPn$ days versus $5.97164\pm 0.00006$ 
from Mayor \& Mazeh and $5.97152\pm0.00002$ from Luyten).  Using our fit value for the B-b 
inclination of $\paramin\pm\errin$, 
Mayor \& Mazeh's value for $K_B=42.1\pm0.3$ km/s, and the Hipparcos parallax, 
one calculates that the semi-major axis of 
component B itself is roughly 760 $\microas$, a value {\em smaller} than 
that found for the astrometric motion.   (This disagreement is most likely 
due to the preliminary nature of the analysis.  However, it may be within the errors of 
parallax measure---Pan, Shao, \& Kulkarni\cite{Pan2004} demonstrated the
Hipparcos distance to Atlas was in error by 10\% or more; a similar error for $\kappa$ Pegasi 
would result in the semi-major axis measurements agreeing within the errors.) 
The rough agreement of the two values suggests that the center-of-light 
of B-b is very close to the location of B itself.  Making the approximation that star b is 
faint, the mass ratio is $M_b/M_B=0.51$.  
S\"oderhjelm's value for the mass of subsystem B-b of $3.13\pm0.2\Msun$ implies that 
component b has mass $1.05\Msun$.  Stars A and B have apparent 
magnitudes $K\approx3.8$ and $K\approx3.6$.  If b is a main-sequence star, 
it would have an apparent magnitude of approximately K=5.8, 
corresponding to a luminosity 7.5 times fainter than B.  
In the scope of this preliminary orbit, this is consistent 
with the proposition that the center-of-light of B-b is located near B.  
The sensitivity of PHASES to lower-mass companions improves for 
longer companion periods.  While the current observations are 
only sensitive to objects of mass $0.1\Msun$ in 6-day orbits about $\kappa$ Pegasi B, 
this drops to 6 Jupiter masses for companions with 5 month periods.

The mean orbital motion of the center-of-light of the B-b subsystem is roughly 900 
$\microas$ per day.  Superimposing the A-Bb orbit adds on average 340 
$\microas$ of motion per day.  During a typical one-hour observation, the total motion 
can be as large as 50 $\microas$.  The motion has little curvature over this time 
period, so the average differential astrometry measure can be taken as the value for
the average time.  Observations with higher precision will require modeling the 
motion within a single night's observing.  

We see no evidence supporting a 4.77-day period companion to $\kappa$ Pegasi A.  The suggested 
amplitude for the velocity curve in Beardsley \& King was roughly 30 km/s, 
corresponding to astrometric motion of star A on order 1.1 mas, an effect that would be seen in 
the PHASES astrometric data if present.  The radial velocity measurement 
precision was of order 1 km/s; the current PHASES data set could detect companions 
at this same level, with masses as small as $0.05 \Msun$ for a 4.77-day period 
(the PHASES uncertainty in $a_{Bb, C.O.L.}$ is $\erran$ $\microas$, 41 
times smaller than the expected 1.1 mas effect and 1.4 times smaller than the astrometric 
perturbation of an object causing a 1 km/s reflex motion).
For companions in longer periods, PHASES is 
more sensitive than these radial velocity observations.  There is no evidence for any 
components other than A, B, and b at this time.

\begin{figure}[!h]
\centerline{\includegraphics[height = 3.75in, angle = 270]{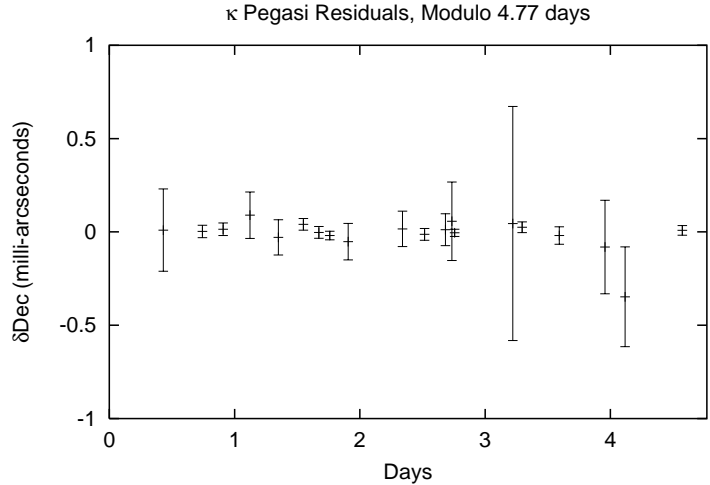}}
\caption{Residuals to the preliminary double-Keplerian 
model for $\kappa$ Pegasi, in differential declination 
(typically our more sensitive dimension), wrapped about 4.77 days.  There is no evidence 
supporting a 4.77 day companion to $\kappa$ Pegasi A, as had been suggested by the 
radial velocity observations of Beardsley \& King.  
For clarity, 6 data points with declination uncertainties larger than the 
semi-major axis of motion of the center-of-light of 
B-b ($\paraman\times10^{-6}$ arcseconds) have not been included in the plot.}
\end{figure}

\section{Conclusions}
The high-precision separation measurements obtained in the PHASES 
survey of sub-arcsecond binaries detect the astrometric 
effect of the unseen companion b in the $\kappa$ Pegasi system.  
The observed astrometric effect is nearly two orders of magnitude larger than the demonstrated 
measurement precision.  The companion star is possibly a main-sequence star of type early G.  
The orbital parameters agree well with the established single-line 
spectroscopic orbit previously identified.  There is no evidence to support the 
suggestion that $\kappa$ Pegasi A has a 4.77 day companion; such a companion would be 
readily identified in the PHASES data.

Phase-referenced differential astrometry is useful for studying low-luminosity companions in S-type 
orbits to binary stars.  This method will allow detection of planetary mass 
companions whose periods are longer than that of $\kappa$ Pegasi b, in sub-arcsecond binaries, should 
they exist.  Such longer-period companions are better suited 
for astrometric than radial velocity observations; even for companions 
in week-long periods our astrometric data rivals the 
sensitivity of previous studies of the $\kappa$ Pegasi system.  
$\kappa$ Pegasi is an ideal testbed for developing astrometric techniques.

\acknowledgments

The PHASES project is funded by the Astronomy Department at the California Institute of 
Technology.  We thank the members of the PTI Collaboration for 
their contributions to the instrument construction, operation, maintenance, and continued 
improvements.  We would like to thank M.~Colavita, K.~Rykoski, and N.~Safizadeh 
for their contributions to this effort.  
Interferometer data was obtained at the Palomar 
Observatory using the NASA Palomar Testbed Interferometer, supported 
by NASA contracts to the Jet Propulsion Laboratory.  
We thank the staff of Palomar Observatory.  M.W.M.~acknowledges 
the support of the Michelson Graduate 
Fellowship program. B.F.L.~acknowledges support from a Pappalardo 
Fellowship in Physics.

\bibliography{muterspaugh}   
\bibliographystyle{spiebib}   

\end{document}